\documentclass[10pt, aps,prx,twocolumn,showpacs,superscriptaddress]{revtex4-2}

\usepackage[colorlinks=true,linkcolor=blue,citecolor=blue,urlcolor=magenta]{hyperref}

\usepackage{amsfonts,mathrsfs,amsmath,amsthm,amssymb}
\usepackage{epsfig}
\usepackage{bm}
\usepackage{array,multirow}
\usepackage{booktabs}
\usepackage{graphicx}
\usepackage{upgreek}
\graphicspath{{images/}}
\usepackage{algorithm}
\usepackage{algpseudocode} 
\usepackage{ulem}
\usepackage{soul}

\newcommand{\la}{\langle}
\newcommand{\ra}{\rangle}

\newcommand{\psik}[1]{|\psi (#1)\ra}

\newcommand{\ifa}{$i$FALQON}

\begin{document}  
\title {\bf 
Imaginary-time-enhanced feedback-based quantum algorithms for ground-state preparation
}

\author{Thanh Nguyen Van Long} 
\affiliation{Faculty of Physics and Engineering Physics, University of Science, Ho Chi Minh City, Vietnam}
\affiliation{Vietnam National University, Ho Chi Minh City 70000, Vietnam}

\author{Lan Nguyen Tran} 
\affiliation{Faculty of Physics and Engineering Physics, University of Science, Ho Chi Minh City, Vietnam}
\affiliation{Vietnam National University, Ho Chi Minh City 70000, Vietnam}

\author{Le Bin Ho} 
\thanks{Electronic address: ho.bin.le.e3@tohoku.ac.jp}
\affiliation{Frontier Research Institute 
for Interdisciplinary Sciences, 
Tohoku University, Sendai 980-8578, Japan}
\affiliation{Department of Applied Physics, 
Graduate School of Engineering, 
Tohoku University, 
Sendai 980-8579, Japan}

\date{\today}

\begin{abstract}
Preparing ground states of strongly correlated quantum systems is a central goal in quantum simulation and optimization. The feedback-based quantum algorithm (FALQON) provides an attractive alternative to variational methods with a fully quantum feedback rule, but it fails in the presence of spectral degeneracies, where the feedback signal collapses and the evolution cannot reach the ground state.
Using the Fermi-Hubbard model on lattices up to \(3\!\times\!3\), we show that this breakdown appears at half-filling on the \(2\!\times\!2\) lattice and extends to both half-filled and doped configurations on the \(3\!\times\!3\) lattice. We then introduce an imaginary-time-enhanced FALQON (\ifa) scheme, which inserts short imaginary-time evolution steps into the feedback loop. This proposed method suppresses excited-state components, escapes degenerate subspaces, restores monotonic energy descent and improves ground-state convergence across the tested fillings. 
We further benchmark its performance against existing methods, showing improved results for \ifa, demonstrating the potential of hybrid real-imaginary-time feedback dynamics for ground-state preparation in strongly correlated quantum systems.
\end{abstract}
\maketitle

\section{Introduction}
Preparing ground states of strongly correlated quantum systems is a central task in quantum chemistry, condensed-matter physics, and quantum information science \cite{RevModPhys.80.1355,RevModPhys.92.015003}. Hybrid quantum-classical approaches, e.g., Variational Quantum Eigensolver (VQE) \cite{Peruzzo2014} and Quantum Approximate Optimization Algorithm (QAOA) \cite{farhi2014quantumapproximateoptimizationalgorithm}, have been widely explored for this purpose. However, their performance is tied to high-dimensional classical parameter optimization and problem-specific ans\"atze, which become increasingly fragile under barren plateaus, noise accumulation, and expressibility constraints \cite{McClean2018,Cerezo2021}. These limitations motivate a different paradigm, i.e., feedback-driven quantum optimization schemes in which all control parameters are generated intrinsically by the quantum dynamics.

The feedback-based algorithm for quantum optimization (FALQON) \cite{PhysRevLett.129.250502} exemplifies this paradigm. Instead of minimizing a classical loss landscape, FALQON directly embeds a Lyapunov-style feedback \cite{doi:10.1049/iet-cta.2009.0508} into the Hamiltonian evolution, 
\(
H(t) = H_p + \beta(t) H_d,
\)
where the control field \(\beta(t)\) is determined from the commutator signal
\(
A(t)=i\langle[H_d,H_p]\rangle_t
\)
to enforce a monotonic decrease of the cost function \(C(t)=\langle H_p\rangle_t\) \cite{doi:10.1049/iet-cta.2009.0508,Hou2019}. 
Recent applications demonstrate that FALQON can efficiently prepare ground states of spin systems, molecular Hamiltonians, and fermionic lattice models in nondegenerate regimes \cite{PhysRevLett.129.250502,PhysRevResearch.6.033336, PhysRevA.109.062603,Abdul_Rahman_2025}.

Compared with variational approaches such as VQE~\cite{Peruzzo2014}, ADAPT-VQE~\cite{Grimsley2019}, or QAOA~\cite{farhi2014quantumapproximateoptimizationalgorithm}, FALQON requires no classical parameter optimization and no problem-specific ans\"atze design, which avoids well-known challenges including barren plateaus, optimizer instability, and limited expressibility~\cite{Cerezo2021}.  Existing studies have largely addressed nondegenerate Hamiltonians, and the behavior of feedback-driven optimization in spectrally degenerate regimes remains unexplored.

In this work, we identify a structural failure of FALQON when applied to systems with degenerate spectra, using the Fermi-Hubbard (FH) model as a controlled testbed \cite{osti_4151988}. Across multiple lattice sizes up to \(3\!\times\!3\), we find that FALQON converges reliably in doped configurations but consistently stalls at half-filling, where particle-hole symmetry generates degenerate manifolds of low-lying states. When the energy trajectory encounters such a manifold, the commutator signal \(A(t)\) becomes strongly suppressed, resulting in feedback amplitudes \(\beta(t)\) that oscillate weakly around zero. These nearly vanishing controls are insufficient to induce further transitions and the energy \(E(t)=\langle H_{p}\rangle_t\) becomes trapped above the true ground state.

To resolve this obstruction, we introduce an imaginary-time-enhanced variant~\cite{McArdle2019,Motta2020}, \ifa{}, in which short imaginary-time evolution steps generated by \(H_p\) are periodically inserted into the feedback loop. Imaginary-time propagation effectively damps excited-state components~\cite{McArdle2019, https://doi.org/10.1002/qute.202100114, Motta2020,PRXQuantum.2.010317,Yeter-Aydeniz2020,Nishi2021,PRXQuantum.3.010320} while preserving the Lyapunov-based descent structure, allowing the dynamics to escape degenerate subspaces and continue toward the ground state. We provide a convergence analysis for the scheme and benchmark its performance on FH lattices up to \(3\times3\), showing monotonic energy descent and high-fidelity ground-state preparation across all fillings within shallow simulation time step. These results demonstrate that \ifa{} provides an effective proof-of-concept framework for improving ground-state convergence in correlated many-body systems.

The structure of the paper is as follows. Section~\ref{sec2} presents the FALQON framework and discusses its limitations. In Sec.~\ref{sec3}, we introduce the \ifa{} framework. Its application to the FH model is demonstrated in Sec.~\ref{sec4}. In Sec.~\ref{sec5}, we compare our approach with other existing frameworks. Section~\ref{sec6} concludes the paper.

\section{FALQON framework}\label{sec2}
Consider a time-dependent Hamiltonian
\begin{equation}
H(t)=H_p+\beta(t)H_d,
\label{eq:H}
\end{equation}
where \(H_p\) is a problem Hamiltonian whose ground state we seek to prepare, \(H_d\) is a driver Hamiltonian, and \(\beta(t)\) is a feedback control updated recursively from the system state, which is constructed to enforce a monotonic decrease in a cost function $C(t)=\la\psi(t)|H_p|\psi(t)\ra = \la H_p\ra_t$. From the Schr{\"o}dinger equation, we derive \cite{PhysRevLett.129.250502}
\begin{equation}
\frac{dC}{dt}=i\,\beta(t)\la[H_d,H_p]\ra_t
=\beta(t)A(t),
\end{equation}
where $A(t)=i\la[H_d,H_p]\ra_t$. By choosing $\beta(t)=-A(t)$, 
we get 
\(\frac{dC}{dt} = -A(t)^2 \le 0,\)
which implies the energy expectation decreases monotonically. The evolution operator is
\begin{equation}
U(T)=\mathcal{T}\exp\!\Big[-i\!\int_0^T\Big(H_p+\beta(t)H_d\Big)dt\Big],
\end{equation}
which, in practice, is implemented via a discrete Trotter expansion as
\begin{align}
U(t)\approx \prod_{k=0}^{T/\Delta t} e^{-iH_p\Delta t} e^{-i\beta_k H_d \Delta t},
\end{align} 
where \(T\) denotes the total evolution time, \(\Delta t\) is the discrete Trotter time step, and $\beta_k=-A_{k-1}$, and 
\begin{align}
|\psi_{k}\rangle = e^{-iH_p\Delta t}e^{-i\beta_{k-1}H_d\Delta t}|\psi_{k-1}\rangle.
\end{align}

In a qubit representation, both \(H_p\) and \(H_d\) can be decomposed into sums of Pauli strings,
i.e., \(
H_p=\sum_\mu c_\mu P_\mu;
H_d=\sum_\nu d_\nu Q_\nu,
\)
where \(P_\mu\) and \(Q_\nu\) denote tensor products of Pauli operators. Consequently, the commutator \([H_d,H_p]\) can also be written as
\begin{equation}
[H_d,H_p]=\sum_\alpha g_\alpha R_\alpha,
\end{equation}
allowing \(\beta_k\) to be estimated on quantum hardware through repeated measurements of the corresponding Pauli observables. In practice, commuting Pauli strings may be grouped into common measurement bases to reduce measurement overhead. The total shot count required to estimate $\beta_k$ with precision $\epsilon$ scales as $N_\text{shots} \lesssim \left(\sum_\alpha |g_\alpha|\right)^2 / \epsilon^2$, which grows polynomially with system size for sparse Hamiltonians such as the FH model. 

FALQON provides a recursive, measurement-driven update rule that avoids classical parameter optimization like VQE or QAOA \cite{Cerezo2021}, with only \(H_d\) and \(\Delta t\) as design inputs. In the ideal continuous-time limit, LaSalle’s invariance principle \cite{doi:10.1137/1.9781611970432} implies that the dynamics may enter an invariant set in which \(\langle[H_d,H_p]\rangle_t = 0\), so that \(\beta(t) = 0\) and the energy stops decreasing \cite{PhysRevResearch.6.033336}.
This typically occurs once the evolution has reached the ground-state manifold, where the feedback signal naturally vanishes \cite{PhysRevLett.129.250502}.

However, in some cases, the system approaches this regime only asymptotically, i.e.,  
\(A(t)\) remains close to zero and may fluctuate around zero at late times as the dynamics approach the invariant set. Consequently, \(\beta(t)\) becomes weak and repeatedly changes sign, leading to damped oscillatory dynamics with little additional energy reduction \cite{PhysRevResearch.6.033336}.
These residual controls are insufficient to produce net energy descent, causing the optimized energy to level off above ground energy. This behavior often arises in half-filled and strong correlation FH lattices, where particle-hole symmetry generates degenerate low-lying spectra with a strongly suppressed commutator response \cite{PhysRevResearch.6.033336}.

\section{Imaginary-time evolution (ITE) enhancement}\label{sec3}
\subsection{\ifa{} protocol}\label{sec3a}
To address the above issue, we incorporate an ITE kick-off scheme. The ITE introduces an effective dissipative dynamics that naturally suppresses excited-state amplitudes and enhances overlap with the ground state \cite{McArdle2019, https://doi.org/10.1002/qute.202100114, Motta2020,PRXQuantum.2.010317,Yeter-Aydeniz2020,Nishi2021,PRXQuantum.3.010320}. The imaginary-time Schr{\"o}dinger equation, \(\partial_\tau \psik{\tau} = -H_p\psik{\tau}\), is analogous to a diffusion equation, where solution gives
\begin{equation}
\psik{\tau} = e^{-H_p\tau} \psik{0} = \sum_n c_n e^{-E_n\tau}|n\rangle,
\end{equation}
where we used $H_p = \sum_n E_n |n\ra\la n|$, and $|\psi(0)\ra = \sum_n c_n |n\ra$.
Here, the higher-energy components decay exponentially faster than the ground state. Taking the limit \(\tau \rightarrow \infty\) gives \(\psik{\tau} \rightarrow c_0 e^{-E_0\tau}|0\rangle\) which converges to its ground state. Applying the Taylor-expand yields
\begin{equation}\label{eq:ite}
   |\Phi_{\Delta\tau}(\psi)\rangle \equiv \psik{\tau + \Delta \tau} \approx \frac{(I - H_p \Delta \tau)\psik{\tau}}{\|(I - H_p \Delta \tau)\psik{\tau}\|}.
\end{equation}

In our \ifa{} protocol, we insert Eq.~(\ref{eq:ite}) periodically between FALQON steps, yielding  
\begin{align}
|\psi_{k}\rangle
= \mathcal{N}\!\left[(I-H_p\Delta\tau)
e^{-iH_p\Delta t}e^{-i\beta_{k-1}H_d\Delta t}|\psi_{k-1}\rangle\right],
\end{align}
where \(\mathcal{N}\) is the normalization. 
The term \((I - \Delta\tau H_p)\) redistributes the population toward lower-energy sectors, reactivating the feedback term \(\beta(t)\) and guiding the evolution back to the true ground state.
This hybrid real-imaginary dynamics unites Lyapunov stability \cite{doi:10.1049/iet-cta.2009.0508,Hou2019} with dissipative convergence toward the energy minimum.

The imaginary-time update
\(
(I-\Delta\tau H_p)
\)
is nonunitary and therefore cannot be directly implemented as a deterministic quantum gate. In this work, we employ this update only at the statevector-simulation level rather than a full hardware-compatible implementation, to investigate its algorithmic effect on the FALQON dynamics. 

For practical quantum-hardware implementations, a hardware-compatible realization could be achieved using variational quantum imaginary-time evolution (VarQITE), where the parameters of a trial state $|\psi(\boldsymbol{\theta})\rangle$ are updated according to McLachlan's variational principle \cite{McArdle2019}. Such implementations generally require additional measurements to estimate the quantum geometric tensor $\mathcal{F}_{ij} = \operatorname{Re}\langle\partial_i\psi|\partial_j\psi\rangle$ and the energy gradient vector $\mathcal{E}_i = -\operatorname{Re}\langle\partial_i\psi|\hat{H}_p|\psi\rangle$, whose costs scale as $\mathcal{O}(N_\text{params}^2)$ and $\mathcal{O}(N_\text{params} \times N_\text{Pauli})$ per update, respectively. 


\subsection{Energy-decrease mechanism of the ITE step}
In this subsection, we analyze how ITE suppresses excited-state components and lowers the energy during the \ifa{} dynamics. Let \(H_p\) be a bounded, positive semidefinite Hamiltonian with spectral norm
\(\|H_p\|=\sup_{|\psi|=1}\langle\psi|H_p|\psi\rangle\le h\).
The hybrid algorithm alternates infinitesimal imaginary-time updates
\(|\Phi_{\Delta\tau}(\psi)\rangle\) with the feedback control defined by
\(\beta(t)=-i\la[H_d,H_p]\ra_t\).
Then, at every ITE update, the normalized cost function 
gives
\begin{align}
   C\big(\Phi_{\Delta\tau}(\psi)\big)
=\frac{\langle (I-\Delta\tau H_p)H_p(I-\Delta\tau H_p)\rangle_\psi}
{\langle (I-\Delta\tau H_p)^2\rangle_\psi},
\end{align}
which obeys the bound
\begin{equation}\label{eq:bound}
C\big(\Phi_{\Delta\tau}(\psi)\big)\le E -
\frac{2\Delta\tau(1-\tfrac{1}{2}h\Delta\tau)}{(1-h\Delta\tau )^2}\mathcal{V},
\end{equation}
for \(0<\Delta\tau \ne \frac{1}{h} < \frac{2}{h}\), where \(E=\la H_p\rangle_\psi,\) 
and the variance \(\mathcal{V}=\langle(H_p-E)^2\rangle_\psi \).
The upper bound implies \(C(\Phi_{\Delta\tau}(\psi))<C(\psi)\) whenever \(\mathcal{V} > 0\). Thus the cost function \(\{C(\Phi_{\Delta\tau}(\psi_k))\}\) decreases until \(\mathcal{V} = 0\).

To derive the bound in Eq.~\eqref{eq:bound}, 
let \(E = \langle H_p \rangle_\psi\) the original cost function, 
\(\mathcal{V} = \langle (H_p - E)^2\rangle_\psi\) its variance,
and recast \(C\big(\Phi_{\Delta\tau}(\psi)\big)
=\frac{T}{D}\), we have 
\begin{align}
   T &= E - 2\Delta\tau\langle H_p^2\rangle_\psi + \Delta\tau^2\langle H_p^3\rangle_\psi,\\
    D &= 1 - 2\Delta\tau E + \Delta\tau^2\langle H_p^2\rangle_\psi. \label{eq:D}
\end{align}
Given 
\( H_p \le E_{\rm max} I\), we have
\begin{align}
  \langle H_p^3\rangle_\psi \le h\langle H_p^2\rangle_\psi,\qquad
\langle H_p^2\rangle_\psi \ge E^2.  
\end{align}
Hence
\begin{align}
 \notag   T &\le E - (2\Delta\tau-\Delta\tau^2 h)\langle H_p^2\rangle_\psi\\
&= E - (2\Delta\tau-\Delta\tau^2 h)(E^2+\mathcal{V}).
\end{align}

For the denominator, using \(\langle H_p^2\rangle_\psi\ge E^2\) gives
\begin{align}
  D \ge 1 - 2\Delta\tau E + \Delta\tau^2 E^2 = (1-\Delta\tau E)^2
\ge (1-\Delta\tau h)^2,  
\end{align}
thus, for \(\Delta\tau \ne \frac{1}{h}\), we have
\(
\frac{1}{D} \le \frac{1}{(1-\Delta\tau h)^2}.
\)
Combining, we get
\begin{align}
    C\big(\Phi_{\Delta\tau}(\psi)\big)
\le \frac{E - (2\Delta\tau-\Delta\tau^2 h)\mathcal{V} - (2\Delta\tau-\Delta\tau^2 h)E^2}
{(1-\Delta\tau h)^2}.
\end{align}

For any \(0<\Delta\tau \ne \frac{1}{h} < \frac{2}{h}\), the coefficient
\((2\Delta\tau-\Delta\tau^2 h)>0\).
Therefore,
\(-(2\Delta\tau-\Delta\tau^2 h)E^2 \le 0\), 
which can be dropped in the numerator yields the clean, conservative bound
\begin{align}
    C\big(\Phi_{\Delta\tau}(\psi)\big)
\le \frac{E}{(1-\Delta\tau h)^2}
- \frac{2\Delta\tau-\Delta\tau^2 h}{(1-\Delta\tau h)^2}\mathcal{V}.
\end{align}

In particular, for sufficiently small \(\Delta\tau\) we obtain the first-order decrease
\begin{align}
    C\big(\Phi_{\Delta\tau}(\psi)\big)
\le E - 2\Delta\tau\mathcal{V} + \mathcal{O}(\Delta\tau^2),
\end{align}
and the rigorous decrease
\begin{align}\label{eq:ite_bound}
    C\big(\Phi_{\Delta\tau}(\psi)\big) \le E -
\frac{2\Delta\tau(1-\tfrac{1}{2}\Delta\tau h)}{(1-\Delta\tau h)^2}\mathcal{V},
\end{align}
for all \(0<\Delta\tau \ne \frac{1}{h} < \frac{2}{h}\) and $\mathcal{V} \ge 0$.

The above analysis indicates that the imaginary-time step locally suppresses excited-state components and decreases the energy whenever the state has nonzero energy variance. Within the \ifa{} dynamics, these periodic nonunitary updates help prevent stagnation near degenerate manifolds where the feedback signal becomes weak, thereby improving the overall convergence behavior observed in the numerical simulations.

\subsection{Spectral origin of the ITE reactivation mechanism}
\label{sec:mechanism}

The previous subsection showed that an ITE step reduces the cost function
for any state with nonzero energy variance. We now examine the spectral
mechanism by which ITE restores the FALQON feedback signal $A(t)$ and
lowers the cost.

Let $\rho(t)=|\psi(t)\rangle\langle\psi(t)|$, such that
$A(t)=i\,\mathrm{Tr}([H_p,\rho(t)]H_d)$. Using
$[H_p,\rho(t)]_{nm}=(E_n-E_m)\rho_{nm}(t)$, we obtain
\begin{equation}
A(t)
=
-2\sum_{n<m}(E_n-E_m)
\,\mathrm{Im}\left[\rho_{nm}(t)(H_d)_{mn}\right].
\label{eq:Aeigen}
\end{equation}
For degenerate eigenstates, $E_n=E_m$, so their contribution to
$A(t)$ vanishes. Thus, if the state is confined to a degenerate
manifold, $A(t)=0$ and the FALQON feedback stops.

To describe the contributions between different energy manifolds, we
group eigenstates with the same energy into $M_j=\{n:E_n=E_j\}$,
with projector $\Pi_j$, and define the inter-manifold coherence
\begin{equation}
g_{jk}(t)
=
\mathrm{Tr}\left[
\Pi_j\rho(t)\Pi_k H_d
\right].
\label{eq:gdef}
\end{equation}
The feedback signal then becomes
\begin{equation}
A(t)
=
-2\sum_{j<k}
(E_j-E_k)\,\mathrm{Im}\,g_{jk}(t).
\label{eq:Adecomp}
\end{equation}

To illustrate the ITE reactivation mechanism, we consider a state
concentrated near a degenerate excited manifold $M_d$, with residual
inter-manifold coherence $g_{cd}(t)$ connecting $M_d$ to a lower-energy
manifold $M_c$ ($E_c<E_d$). Since
\(
K=I-\Delta\tau H_p
\)
acts uniformly within each energy manifold,
\begin{equation}
K\Pi_j=k_j\Pi_j,
\;
k_j=1-\Delta\tau E_j.
\end{equation}
The normalized ITE update $\rho\rightarrow K\rho K/D$ then gives
\begin{equation}
g'_{cd}
=
\frac{k_c k_d}{D}g_{cd}.
\label{eq:gainrule}
\end{equation}
Near $M_d$, $D\simeq k_d^2$, and hence
\begin{equation}
g'_{cd}
\simeq
\frac{k_c}{k_d}g_{cd}
=
\frac{1-\Delta\tau E_c}
     {1-\Delta\tau E_d}g_{cd}.
\label{eq:gainlimit}
\end{equation}
For $E_c<E_d$ and $1-\Delta\tau E_d>0$, the prefactor larger than one,
so ITE amplifies $|\mathrm{Im}\,g_{cd}|$. Through
Eq.~\eqref{eq:Adecomp}, this enhancement restores the feedback signal and resumes the FALQON energy descent. Equation~\eqref{eq:gainlimit} also
shows that the reweighting becomes stronger as $\Delta\tau$ increases
within the small-step regime, $\Delta\tau E_d<1$.

We emphasize that this mechanism differs from standalone
QITE~\cite{Motta2020,McArdle2019}, which does not rely on the feedback
signal $A(t)$. In \ifa{}, ITE instead reweights the spectral components
that contribute to the FALQON feedback. A direct comparison with
standalone QITE is presented in the next section.

\section{Demonstration on Fermi-Hubbard model}\label{sec4}
\subsection{Model}
The Fermi-Hubbard (FH) model describes interacting fermions on a lattice and captures the competition between kinetic hopping and on-site repulsion. It provides a minimal framework for phenomena such as the Mott transition, antiferromagnetism, and unconventional superconductivity. Its Hamiltonian is
\begin{equation}
H_{\rm FH} = -J \sum_{\langle i,j \rangle, \sigma}
\left( c^\dagger_{i\sigma} c_{j\sigma} + c^\dagger_{j\sigma} c_{i\sigma} \right)
+ U \sum_{i} n_{i\uparrow} n_{i\downarrow},
\label{eq:fh_ham}
\end{equation}
where \(J\) denotes the hopping amplitude between nearest-neighbor sites \(\langle i,j \rangle\), and \(U\) represents the on-site Coulomb repulsion between fermions of opposite spin.
The fermionic creation and annihilation operators \(c^\dagger_{i\sigma}\) and \(c_{i\sigma}\) obey the canonical anticommutation relations
\(\{c_{i\sigma}, c_{j\sigma'}^\dagger\} = \delta_{ij}\delta_{\sigma\sigma'}\),
and the local number operator is \(n_{i\sigma} = c^\dagger_{i\sigma} c_{i\sigma}\).

For a lattice with \(L\) sites and particle numbers \(N_\uparrow\) and \(N_\downarrow\), the filling is
\(\nu = N/L\) with \(N=N_\uparrow+N_\downarrow\).
Half-filling \((\nu=1)\) corresponds to one particle per site on average and typically exhibits strong correlations and insulating behavior at large \(U/J\). In this regime, particle-hole symmetry leads to degenerate low-lying spectra.
Away from half-filling \((\nu\neq 1)\), the system is doped, particle-hole symmetry is broken, and mobile carriers appear, often restoring metallic behavior. In simulations, configurations are specified by the spin populations, e.g., \((1,2)\) for a three-particle doped case or \((2,2)\) for the half-filled four-site lattice.

This model captures the essential interplay between kinetic hopping and on-site interaction, making it a prototypical system for studying strongly correlated electrons \cite{Izyumov_1995,RevModPhys.68.13}, Mott insulators \cite{RevModPhys.78.17,PhysRevB.82.075106,PhysRevLett.134.016503,PhysRevResearch.5.043267}, and magnetism~\cite{Gall2021,Chen2025}. Despite its simplicity, the model exhibits a rich variety of complex quantum materials. Beyond basic magnetic ordering like ferromagnetism and various forms of antiferromagnetism, it is known to host exotic phases including charge-density waves (CDWs)~\cite{PhysRevB.106.L081107,vanEfferen2021}, electronic liquid-crystal states with broken rotational symmetry~\cite{annurev:/content/journals/10.1146/annurev-conmatphys-070909-103925,PhysRevB.64.195109,RevModPhys.84.497}, and spin-liquid phases in frustrated regimes~\cite{Savary_2017,RevModPhys.89.025003}. It also provides a central microscopic framework for unconventional superconductivity \cite{annurev:/content/journals/10.1146/annurev-conmatphys-031620-102024}. 

\begin{figure}[t]
    \centering
    \includegraphics[width=\linewidth]{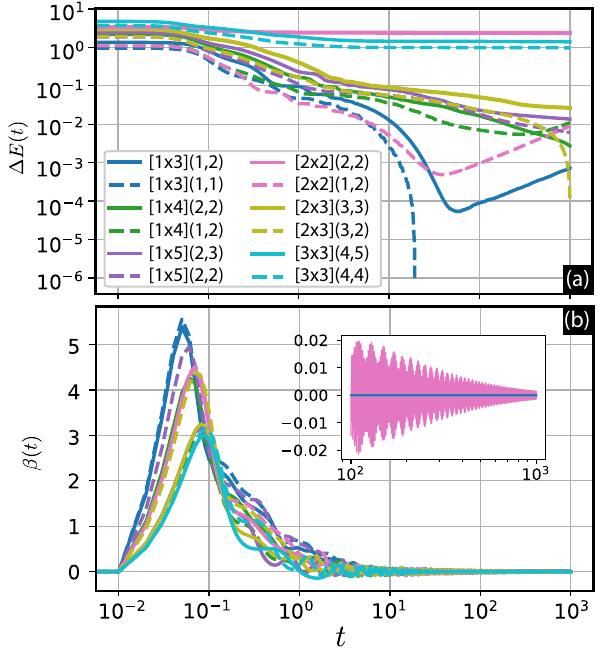}
    \caption{\textbf{Dynamics of the FALQON without ITE.}
(a) Time evolution of the energy difference \(\Delta E(t)=E(t)-E_0\) for various FH lattices and fillings.
Doped systems \(([1\!\times\!3]\), \([1\!\times\!4]\), \([2\!\times\!3])\) exhibit smooth exponential convergence toward the exact ground energy, while half-filled configurations \(([2\times2]\), \([3\times3])\) stagnate at finite energy errors.
(b) Corresponding feedback amplitude \(\beta(t)\) as defined by the Lyapunov control law \(\beta(t)=-A(t)\).
The early peak in \(\beta(t)\) accelerates energy descent, but in half-filled lattices the feedback collapses to zero prematurely, halting further optimization.
The inset shows long-time oscillations of \(\beta(t)\) for the \([2\!\times\!2]\) and \([3\!\times\!3]\) cases, demonstrating symmetric fluctuations around zero and confirming feedback quenching due to degeneracy.
Together, these results reveal that pure FALQON effectively prepares ground states only for non-degenerate systems, motivating the introduction of an imaginary-time enhancement. In the legend, \([i\!\times\!j]\) stands for the lattice site and \((k,l)\) stands for \((N_\uparrow, N_\downarrow )\).}
    \label{fig1}
\end{figure}

\begin{figure}[t]
    \centering
    \includegraphics[width=\linewidth]{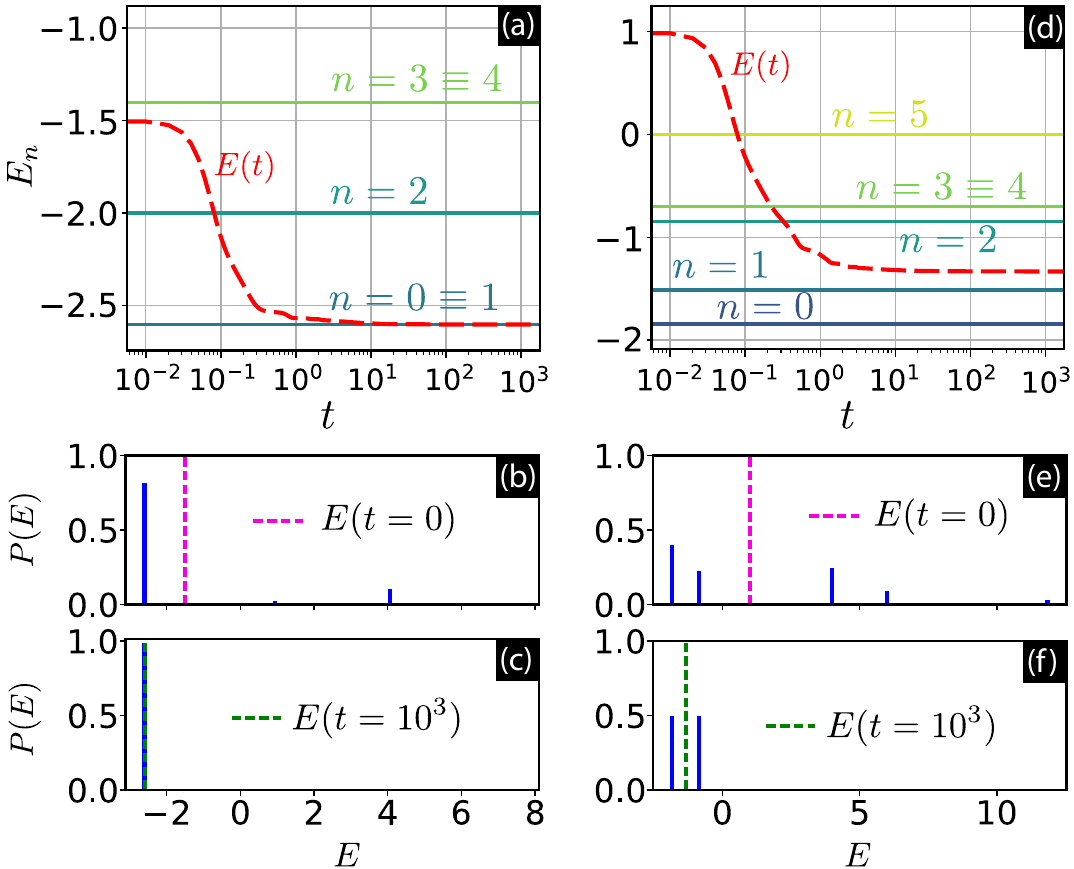}
    \caption{\textbf{Energy spectra and population dynamics for the doped and half-filled \([2\!\times\!2]\) FH systems under FALQON without ITE.} (a,d) Mean energy \(E(t)\) (dashed red) overlaid on the energy spectra \(E_n\) of \(H_{\rm FH}\).
    In the doped case (a), \(E(t)\) descends toward the ground state without crossing any degenerate levels.
    In the half-filled case (d), \(E(t)\) crosses a manifold of degenerate excited states, suppressing the feedback response and halting further relaxation.
    (b, c) Probability distributions \(P(E)\) at \(t=0\) and \(t=10^3\) for the doped system, showing collapse into the ground state. (e, f) Corresponding distributions for the half-filled system, where significant weight remains in low-lying excited states due to trapping within a subspace.}
    \label{fig2}
\end{figure}

To numerically simulate this system on quantum computers, \(H_{\rm FH}\) needs to be formulated in Pauli strings. For an \(N\) sites lattice, we use Jordan-Wigner (JW) mapping to encodes the \(2N\) spin orbitals to \(2N\) qubits~\cite{Jordan1928}. This transformation maps local fermionic operators to non-local Pauli strings involving chains of $Z$ operators to preserve the anti-commutation relations. Consequently, the hopping terms in the Hamiltonian are transformed into multi-qubit strings, while on-site interactions are mapped to local diagonal terms. The explicit derivation and the full form of the mapped Hamiltonian are provided in the Appendix. 

After mapping the fermions to qubits, we simulate ground-state preparation directly on quantum registers. We focus on the strongly correlated regime $U/J = 5.0$, which lies deep in the Mott insulating phase \cite{RevModPhys.78.17,PhysRevB.82.075106,PhysRevLett.134.016503,PhysRevResearch.5.043267}. In this regime, charge fluctuations are suppressed and spin correlations dominate, generating strong entanglement that beyond the reach of mean-field methods, tensor-network truncations, and Monte Carlo techniques~\cite{SCHOLLWOCK201196,Orus2019,PhysRevLett.94.170201}.

Our benchmarking area includes linear 1D lattices \(\big([1\!\times\!3]\), \([1\!\times\!4],\) and \([1\!\times\!5]\big)\), and square 2D lattices \(\big([2\!\times\!2]\), \([2\!\times\!3],\) and \([3\!\times\!3]\big)\). 
Hereafter, \([i\!\times\!j]\) labels the lattice size, and \((k,l)\) indicates the number of spin-up and spin-down fermions, \(N_\uparrow\) and \(N_\downarrow\), respectively.
For example, the largest \([3\!\times\!3](4,5)\) lattice corresponds to \(9\) lattice sites and \(18\) spin orbitals, requiring \(18\) qubits after the JW transformation.
All simulations were performed using the Qiskit statevector simulator, providing a noiseless benchmark of algorithmic performance without decoherence, sampling noise, or thermal fluctuations. Therefore, the results here interpret as proof-of-concept demonstrations of the \ifa{} framework. We set \(H_p = H_{\rm FH}\) and use the hopping term \(H_d = -J\sum_{\langle i,j\rangle,\sigma}(c^\dagger_{i\sigma}c_{j\sigma}+h.c.)\), which conserves particle number while not commuting with \(H_p\), thereby enabling effective state transitions.

\subsection{FALQON without ITE}  
Figure~\ref{fig1} shows the dynamical behavior of FALQON without ITE, evaluated various FH lattice geometries and fillings.
Fig.~\ref{fig1}(a) plots the time evolution of the energy difference \(\Delta E(t)=E(t)-E_0\), where \(E_0\) is the exact ground-state energy obtained from exact diagonalization of $H_{\rm FH}$, and \(E(t) = \la H_{\rm FH}\ra_t\).

For small and moderately asymmetric lattices such as \([1\!\times\!3], [1\!\times\!4], [1\!\times\!5]\), and \([2\!\times\!3]\), \(\Delta E(t)\) decays exponentially over several orders of magnitude, reaching values below \(10^{-2}\) at $t = 10^3$.
This monotonic decay confirms that FALQON efficiently drives the system toward its true ground state through the self-consistent feedback law \(\beta(t)=-A(t)\).

For the \([2\!\times\!2]\) lattice, the behavior depends strongly on the filling configuration. In the doped configuration \([2\!\times\!2]\ (1,2)\), the particle-hole symmetry breaks, and thus the convergence remains exponential, similar to asymmetric lattices. In contrast, the half-filled case \([2\!\times\!2]\ (2,2)\) shows a plateau of \(\Delta E(t)\) above \(\!10^{0}\). This behavior indicates that the evolution has entered an invariant subspace in which the expectation value of the commutator \(A(t)=i\langle[H_d,H_p]\rangle_t\) averages to zero [see the inset Fig.~\ref{fig1}(b)].

For the \([3\!\times\!3]\) lattice, both doped and half-filled configurations show a similar convergence plateau. After an initial transient decay, \(\Delta E(t)\) flattens into a broad plateau as the feedback amplitude diminishes, and the system remains stuck above the ground-state manifold.

Figure~\ref{fig1}(b) displays the time evolution of the feedback field \(\beta(t)\). The signal initially rises in response to large values of the commutator \(A(t)\), and subsequently decreases as the system approaches a stationary regime. In trajectories that successfully converge to the ground-state energy, \(\beta(t)\) smoothly decays to zero. By contrast, in non-convergent cases where the dynamics become trapped in a local minimum, \(\beta(t)\) settles into small oscillations around zero. The inset compares \(\beta(t)\) for the \([2\!\times\!2] (2,2)\) and \([1\!\times\!3] (1,2)\) configurations over \(t\in[10^{2},10^{3}]\), showing that the former exhibits rapidly damped oscillations about zero, whereas the latter displays a monotonic decay toward zero.

To further elucidate the behavior in Fig.~\ref{fig1}, we examine the energy spectrum \(\{E_n\}\) of $H_{\rm FH}$, i.e., \(H_{\rm FH} = \sum_nE_n |n\ra\la n|\), and the populations \(P(E) = |\la n|\psi(t)\ra|^2\) for the \([2\!\times\!2]\) doped and half-filled systems. Figures~\ref{fig2}(a-c) correspond to the doped configuration \((1, 2)\), while Figs.~\ref{fig2}(d-f) show the half-filled case \((2,2)\). 

In the doped case shown in Fig.~\ref{fig2}(a), the low-lying spectrum exhibits several two-fold degeneracies, including \(E_0 = E_1\) and \(E_3 = E_4\).
Remarkably, the trajectory of the mean energy \(E(t) = \la H_{\rm FH}\ra_t\) decreases monotonically toward the ground-state manifold without crossing any of these degenerate levels except \(E_0 = E_1 =\) ground energy. The absence of degeneracy crossings ensures that \(A(t)\) remains finite throughout the evolution, enabling the feedback field to continue directing population toward lower energies.
As shown in Figs.~\ref{fig2}(b, c), the initially broad energy distribution \(P(E)\) rapidly collapses into a sharply peaked ground-state population by \(t=10^3\). This behavior aligns with the exponential decay of \(\Delta E(t)\) in Fig.~\ref{fig1}(a) and demonstrates that FALQON achieves reliable ground-state preparation in the doped configuration.

In contrast, the half-filled system shown in Fig.~\ref{fig2}(d) exhibits a qualitatively different mechanism. Here, the trajectory \(E(t)\) encounters a manifold of degenerate excited states, such as the pair \(E_{3}=E_{4}\). Once \(E(t)\) enters this degenerate sector, \(A(t)\) becomes strongly suppressed, and the resulting feedback field \(\beta(t)\) decays toward zero with only weak residual oscillations. These small controls are insufficient to induce transitions out of the degenerate manifold, causing the energy evolution to trap above the true ground energy. The probability distributions in Figs.~\ref{fig2}(e, f) further confirm this behavior, where even at long times, the population remains distributed over several low-lying excited states rather than concentrating at \(E_0\). Thus, in the half-filled configuration, the trajectory becomes trapped in a local subspace after crossing a degenerate manifold, preventing full relaxation to the ground state.

\subsection{\ifa}  
Figure~\ref{fig3} shows the energy difference \(\Delta E(t)=E(t)-E_0\) for all tested lattices under the \ifa{} protocol. An imaginary-time step of \(\Delta\tau=0.05\) is applied after every two feedback layers. Unlike the behavior in Fig.~\ref{fig1}, all trajectories now exhibit a clean monotonic exponential decay. Systems from one-dimensional chains \([1\!\times\!3]-[1\!\times\!5]\) to two-dimensional lattices \([2\!\times\!2], [2\!\times\!3], [3\!\times\!3]\) converge to the exact ground-state energy with final errors below \(10^{-5}\). Crucially, the \([2\!\times\!2](2,2)\) and \([3\!\times\!3](k, l)\) cases, both of which stalled under pure FALQON, now follow the same smooth descent as the other configurations. The ITE step accelerates convergence by continuously shifting amplitude toward lower energies, shortening the relaxation time. This result shows that hybrid real-imaginary-time feedback improves ground-state convergence, making \ifa{} as a robust framework for strongly correlated quantum systems.

\subsection{\([3\!\times\!3]\) lattice FH model}
This is the most strongly correlated system considered in this work, and its convergence behavior under pure FALQON remains poor for both doped and half-filled configurations, as shown in the main text. To clarify the origin of this failure, Fig.~\ref{fig4} displays the probability distributions over the energy eigenbasis at the beginning and end of the FALQON evolution.

The upper panels show the initial energy distributions for the doped (a) and half-filled (b) configurations. In both cases, the initial state carries substantial weight across high-energy excited levels, with dominant support well above the ground energy (red dashed line). Such broad high-energy populations require the FALQON dynamics to transfer probability across many levels, which becomes increasingly difficult on larger correlated lattices.

\begin{figure}[t]
    \centering
   \includegraphics[width=\linewidth]{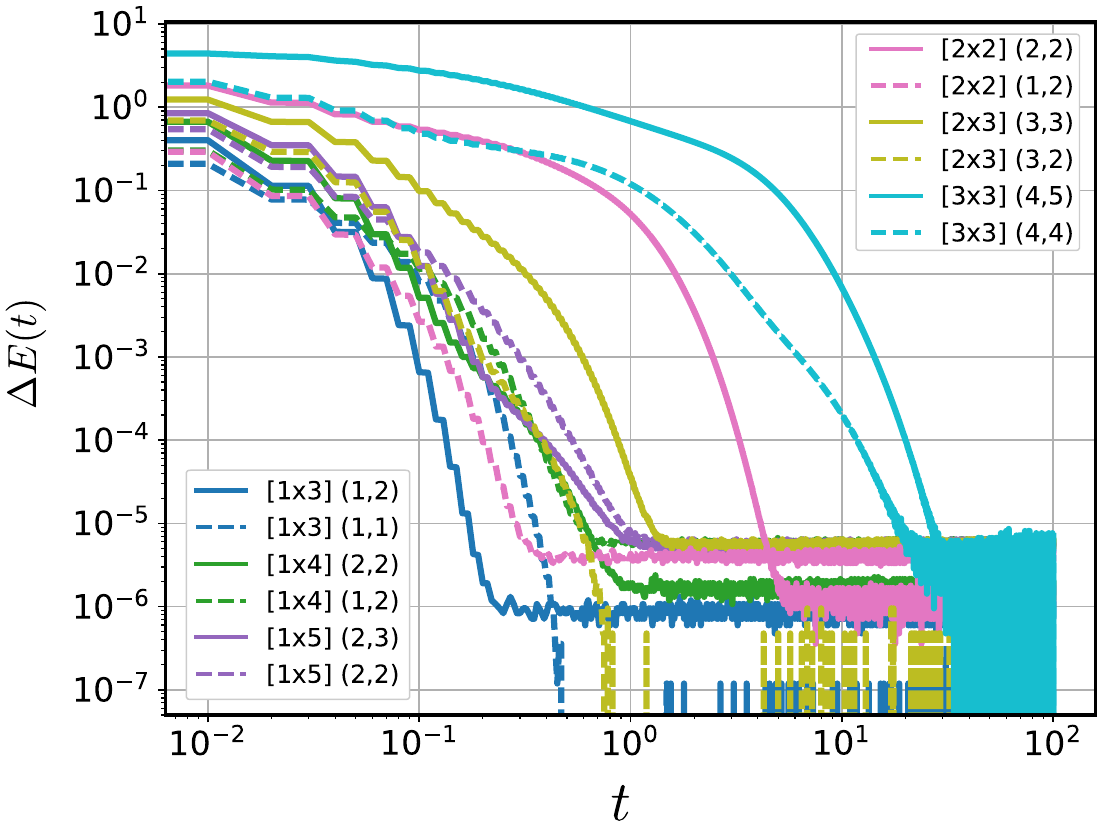}
    \caption{\textbf{Energy convergence of the \ifa{}.}
    Time evolution of the energy difference \(\Delta E(t)=E(t)-E_0\) for various FH lattices.
    An imaginary-time step of \(\Delta\tau=0.05\) is inserted after every two feedback layers.
    All configurations now converge monotonically toward the exact ground energy, reaching relative errors below \(10^{-5}\).
    The exponential decay across more than six decades confirms that \ifa{} restores stable ground-state preparation.}
    \label{fig3}
\end{figure}

\begin{figure}[t]  
    \centering
\includegraphics[width=\linewidth]{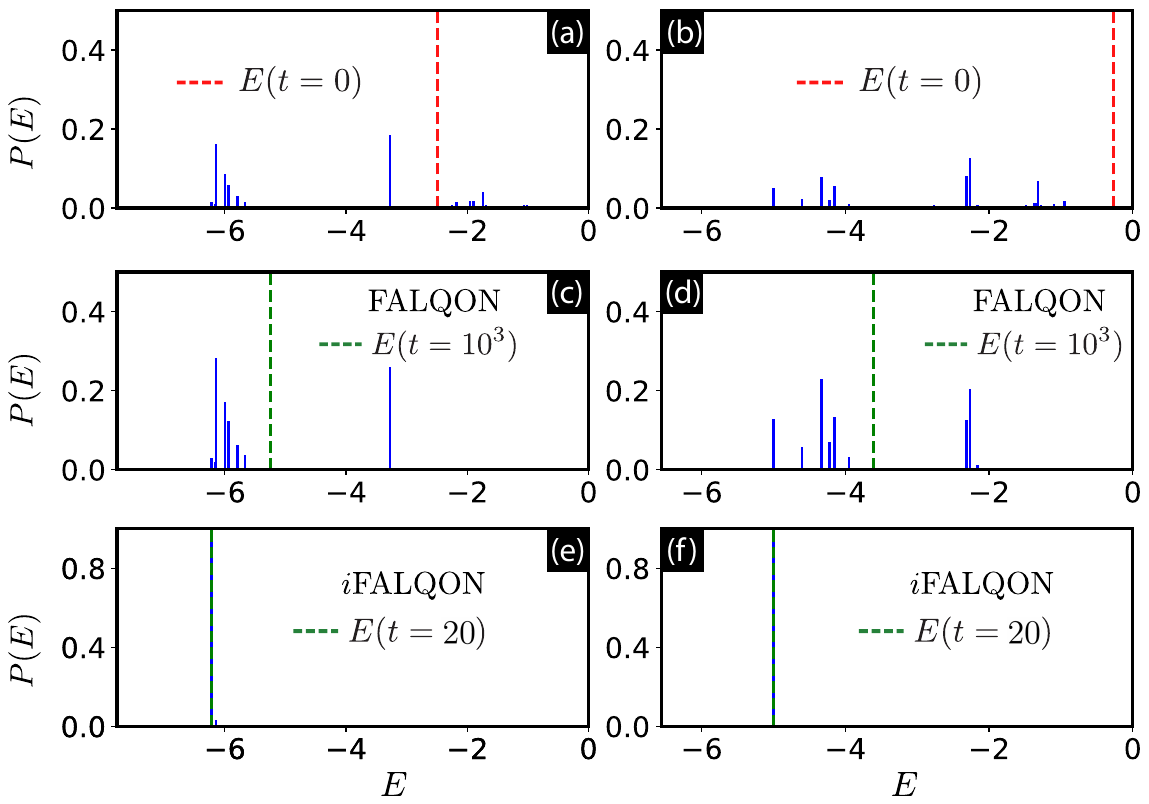}
    \caption{\textbf{Energy distributions for the \([3\!\times\!3]\) FH lattice}. (a, b) Initial energy populations for doped and half-filled configurations, showing dominant support at high excited levels above the ground state (red dashed line). (c, d) Final distributions under pure FALQON remain spread over excited states, indicating failure to reach the ground state. (e, f) \ifa{} collapses the distributions onto the ground state, demonstrating successful convergence.}
    \label{fig4}
\end{figure}

The middle panels (c, d) show the final energy distributions after a long FALQON evolution. In contrast to smaller lattices discussed in the main text, the \([3\!\times\!3]\) system fails to concentrate probability near the ground state. Although some redistribution occurs, substantial weight remains on excited levels and the dominant peaks lie far above the ground energy (green dashed line). This persistent high-energy support matches the plateau in \(\Delta E(t)\) and indicates that the feedback updates cannot drive transitions across the densely packed low-lying spectrum of the \([3\!\times\!3]\) lattice.

The bottom panels (e, f) display the final distributions under \ifa{}. In this case, both configurations successfully converge to the ground state, and the transition occurs noticeably faster than in the pure FALQON protocol.

These distributions show that, on the \([3\!\times\!3]\) lattice, pure FALQON is unable to drive the wave function into the low-energy sector. However, introducing periodic imaginary-time steps (\ifa{}) suppresses high-energy amplitudes and enables population flow toward the ground state. 

\begin{figure}[b]  
    \centering
\includegraphics[width=\linewidth]{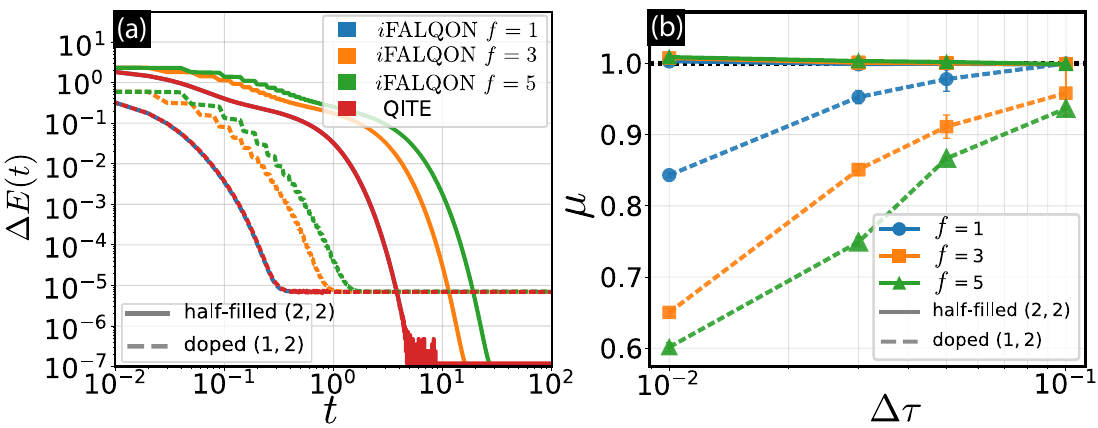}
  \caption{\textbf{Comparison of \ifa{} and standalone QITE on the
$[2\times2]$ FH lattice.}
(a) Energy error $\Delta E(t)=E(t)-E_0$ for the half-filled $(2,2)$
(solid lines) and doped $(1,2)$ (dashed lines) configurations at
$\Delta\tau=0.05$. \ifa{} is shown for ITE insertion periods
$f=1,3,$ and $5$, while standalone QITE applies an ITE update at every
step. 
(b) Ratio $\mu=N_{\rm ITE}^{i\rm FALQON}/N_{\rm ITE}^{\rm QITE}$
of the number of ITE updates required to reach $\Delta E<10^{-5}$ as a
function of $\Delta\tau$.}
    \label{fig5}
\end{figure}

\subsection{Comparison with standalone imaginary-time evolution}
We next compare \ifa{} with standalone QITE~\cite{Motta2020,McArdle2019}. Both methods start
from the same initial state and use the same first-order imaginary-time
update. 
We define the ITE
insertion period $f$ is the number of FALQON steps between two time
ITE updates, with $f=1$ corresponding to an ITE update after every
FALQON step.

Figure~\ref{fig5}(a) shows the energy error
$\Delta E(t)=E(t)-E_0$ for the doped $[2\times2](1,2)$ and half-filled
$[2\times2](2,2)$ configurations at $\Delta\tau=0.05$. 
In standalone QITE, each evolution step is an ITE update, while for
\ifa{}, we consider insertion periods $f=1,3,$ and $5$.
Increasing $f$ slows the convergence because ITE updates are applied
less frequently.

We next examine how the ITE insertion period affects the number of ITE
updates required to reach a given accuracy. To quantify this effect,
Fig.~\ref{fig5}(b) shows
\begin{equation}
\mu=
\frac{N_{\rm ITE}^{i\rm FALQON}}
     {N_{\rm ITE}^{\rm QITE}},
\end{equation}
where $N_{\rm ITE}$ is the number of ITE updates required to reach
$\Delta E<10^{-5}$. Thus, $\mu<1$ means that \ifa{} requires fewer
ITE updates than standalone QITE. 

For the doped configuration, $\mu$
decreases with increasing $f$, reaching approximately $0.60$ at
$f=5$ and $\Delta\tau=0.01$. In contrast, $\mu$ remains close to
one at half filling over the tested range of $f$ and $\Delta\tau$.
Thus, the FALQON evolution reduces the required number of ITE updates
when the feedback remains active, as shown in the doped case. However, there is
no such reduction when the feedback is suppressed at half filling.

A reduction in the number of ITE updates does not imply a lower total computational cost, since \ifa{} also involves FALQON evolution. We estimate the costs as
\begin{align}
\mathcal{C}_{i\rm FALQON}
&=
N_{\rm ITE}^{i\rm FALQON}\mathcal{C}_{\rm ITE}
+
N_{\rm F}\mathcal{C}_{\rm F}, \notag\\
\mathcal{C}_{\rm QITE}
&=
N_{\rm ITE}^{\rm QITE}\mathcal{C}_{\rm ITE},
\end{align}
where $\mathcal{C}_{\rm ITE}$ and $\mathcal{C}_{\rm F}$ denote the
costs of one ITE and FALQON step, respectively. The equal-cost
condition, $\mathcal{C}_{i\rm FALQON}=\mathcal{C}_{\rm QITE}$, is
\begin{equation}
N_{\rm F}\mathcal{C}_{\rm F}
=
(1-\mu)\mathcal{C}_{\rm QITE}.
\label{eq:cost_condition}
\end{equation}
Thus, the two methods have the same total cost only when the FALQON cost
equals the cost saved by fewer ITE updates. 

As discussed in Sec.~\ref{sec3a}, a QITE implementation requires
measurements of the quantum geometric tensor $\mathcal{F}_{ij}$ and
energy gradient $\mathcal{E}_i$, with the costs scaling as
$\mathcal{O}(N_{\rm params}^2)$ and
$\mathcal{O}(N_{\rm params}N_{\rm Pauli})$, respectively, per update.
In \ifa{}, the FALQON step instead evaluates the feedback signal
\(
A(t)
\)
which requires $\mathcal{O}(N_{\rm Pauli})$, where
$N_{\rm Pauli}$ is the number of Pauli terms in $i[H_d,H_p]$. 
A hardware-level comparison accounts for the circuit and measurement costs of the FALQON steps and depends on the specific ITE implementation, which is beyond our
statevector simulations.

\section{Benchmarking}\label{sec5}
In this section, we compare the performance of the \ifa{} model with other existing approaches, including QAOA and VQE.
\subsection{FH model}
Figure~\ref{fig6} summarizes the benchmarking results for ground-state energy estimation of the FH model across different lattice sizes for various frameworks, including QAOA, VQE, FALQON, and \ifa{}.
Figure~\ref{fig6}(a) corresponds to the doped regime, while Fig.~\ref{fig6}(b) shows the half-filling case.
The energy error is defined as $\Delta E = E - E_0$, where $E_0$ denotes the exact ground-state energy and $E$ denotes the final energy at the last iteration.

QAOA and VQE are both hybrid quantum-classical algorithms based on unitary variational ans\"atze optimized through a classical feedback loop.
In QAOA, the trial state is constructed by applying $p=8$ alternating layers of time evolution under the problem Hamiltonian $H_c$ and a non-commuting mixer Hamiltonian $H_d$. We adopt the same Hamiltonian formulations for $H_c$ and $H_d$ as used in FALQON. The parameters are optimized using the COBYLA optimizer over $5000$ iterations.

VQE employs a parameterized quantum circuit to minimize the energy expectation value of the Hamiltonian. We use the Efficient Hamiltonian Variational (EHV) ansatz to reduce circuit depth arising from the JW transformation. The simulation is executed with $p=5$ layers, where optimization process consists of $2000$ COBYLA steps.

These limitations are clearly reflected in Fig.~\ref{fig6}.
In both the doped (a) and half-filled (b) regimes, QAOA and VQE exhibit relatively large energy errors that vary weakly with lattice size.
In particular, QAOA shows pronounced fluctuations, indicating difficulty in maintaining consistent convergence across different lattice geometries.
VQE yields more stable results but remains confined to an error floor set by the expressibility of the ansatz and the optimization process.

The original FALQON protocol improves upon variational methods by replacing global optimization with feedback-based unitary evolution.
In the doped regime, FALQON achieves moderate error reduction for certain lattice sizes but displays noticeable performance variability.
At half filling, where the ground state exhibits stronger correlations and enhanced symmetry constraints, the purely unitary feedback dynamics become less effective, leading to a degradation in accuracy for larger lattices.

\begin{figure}[t]  
    \centering
\includegraphics[width=\linewidth]{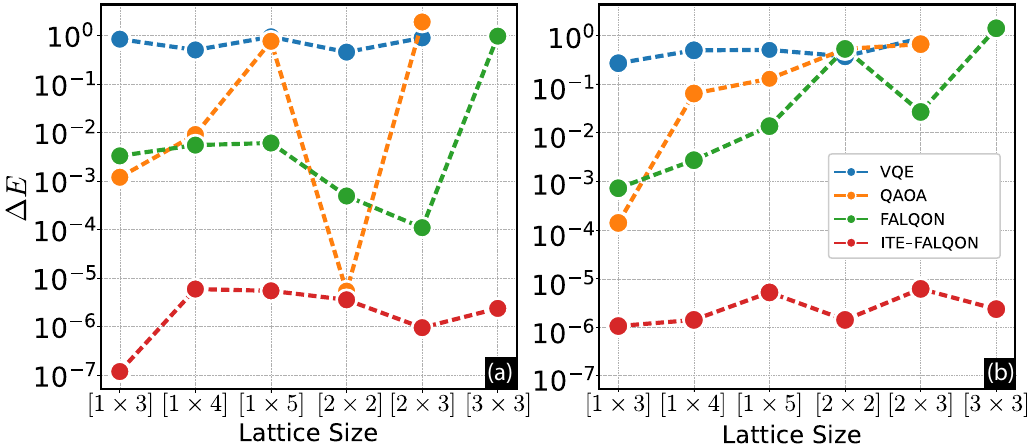}
    \caption{\textbf{Benchmarking between QAOA, VQE, FALQON, and \ifa{}}: Energy error $\Delta E = E - E_0$ as a function of lattice size for the FH model. (a) The doped regime,  and (b) the half-filling case.
    Results are shown for VQE (EHV ansatz), QAOA, FALQON, and the proposed \ifa{} protocol.}
    \label{fig6}
\end{figure}

In contrast, the \ifa{} framework achieves lower energy errors than the compared methods for the tested systems.
By incorporating imaginary-time evolution into the feedback protocol, excited-state components are exponentially suppressed, enabling rapid convergence toward the ground state.
As shown in Fig.~\ref{fig6}, the resulting energy error remains at the level of $10^{-6}$-$10^{-7}$ across all lattice sizes, with only weak dependence on filling fraction or system geometry.

The comparison between (a) and (b) further highlights the robustness of \ifa{}.
While variational and unitary methods exhibit increased sensitivity at half filling due to stronger many-body correlations, \ifa{} maintains uniform performance.
This demonstrates that nonunitary, feedback-assisted evolution  provides a robustness for
ground-state preparation in strongly correlated fermionic systems.

\subsection{Ground-state potential energy of hydrogen chains}

To assess the performance of the \ifa{} on molecular systems, we benchmark the ground-state potential energy surfaces of hydrogen chains, ${\rm H}_4$ and ${\rm H}_6$, in the STO-3G basis. The interatomic distance $r$ between nearest-neighbor hydrogen atoms is varied from $0.4$~\AA{} to $2.8$~\AA{}, spanning both the weakly correlated equilibrium region and the strongly correlated stretched regime. These systems provide standard yet challenging testbeds for evaluating the robustness of ground-state solvers in the presence of electron correlation.

The performance of \ifa{} is compared against the VQE and the full configuration interaction (FCI) reference. For the VQE simulations, we employ the unitary coupled-cluster ansatz with single and double excitations (UCCSD) \cite{RevModPhys.92.015003,Bauer2020}, with parameters optimized using the Adam optimizer at a learning rate of $0.1$ for $200$ iterations (for H$_4$) and $1000$ iterations (for H$_6$) at each molecular geometry. For \ifa{}, the simulations are carried out with a total evolution time $T=100$, a time step $dt=0.005$, and an imaginary-time increment $\Delta\tau=0.15$ applied every two FALQON steps.

As shown in Fig.~\ref{fig7}(a, b), \ifa{} reproduces the FCI ground-state potential energy curves of the hydrogen chains ${\rm H}_4$ and ${\rm H}_6$ with high accuracy over the full range of interatomic distances $r$. The agreement holds both near equilibrium and deep in the stretched regime, where strong electronic correlations become significant. In contrast, the VQE-UCCSD exhibits noticeable deviations from the FCI curve as $r$ increases, particularly for ${\rm H}_6$, where the variational optimization becomes increasingly unstable and sensitive to the choice of ansatz and optimizer \cite{Burton2023,Fan2023,Bauer2020}.

The corresponding energy deviations relative to FCI, displayed in Fig.~\ref{fig7}(c,d), further demonstrate the robustness of \ifa{}, which maintains small and smooth errors across all bond lengths for both systems. In contrast, VQE-UCCSD exhibits increasingly large and nonuniform deviations as $r$ increases, most prominently for ${\rm H}_6$, indicating reduced stability in the strongly correlated regime.

These results illustrate the advantage of \ifa{}’s feedback-assisted imaginary-time evolution, which mitigates optimization instabilities and avoids the sensitivity to ansatz choice and optimizer settings commonly encountered in variational approaches \cite{Burton2023,Fan2023,Bauer2020}. Compared with standalone QITE approaches, \ifa{} retains the feedback-driven dynamics of FALQON while applying imaginary-time updates only periodically, potentially reducing the number of costly imaginary-time optimization steps while still improving convergence. As system size and correlation strength grow, \ifa{} maintains stable convergence and reliable energy estimates, highlighting its effectiveness for near-term quantum simulations of strongly correlated molecular systems.

\begin{figure}[t]
    \centering
    \includegraphics[width=\linewidth]{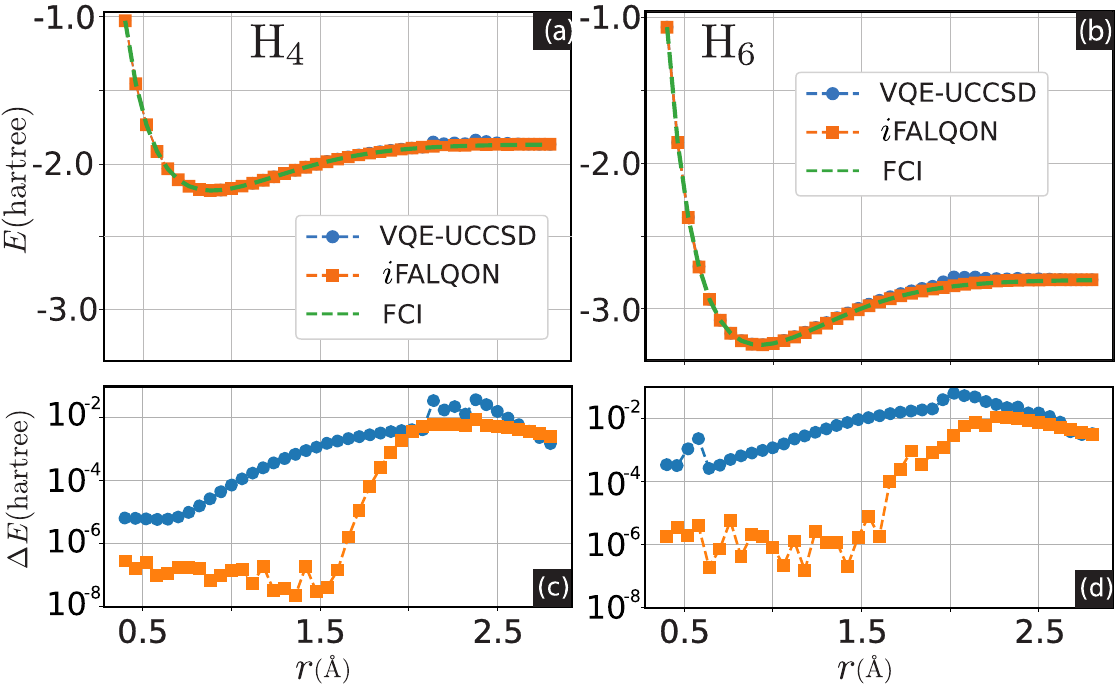}
    \caption{\textbf{\ifa{} accurately reproduces the ground-state potential energy surfaces of hydrogen chains.} Ground-state potential energy of the hydrogen chains ${\rm H}_4$ (a) and ${\rm H}_6$ (b) as a function of the interatomic distance $r$, obtained using \ifa{}, VQE-UCCSD, and FCI. The corresponding energy errors relative to FCI are shown in (c) for ${\rm H}_4$ and in (d) for ${\rm H}_6$.}
    \label{fig7}
\end{figure}

\section{Conclusion}\label{sec6}
We have shown that FALQON converges reliably only when the energy trajectory avoids spectral degeneracies. If the Hamiltonian contains degenerate eigenlevels and the optimal descent path crosses such a manifold, the feedback signal becomes strongly oscillate and the dynamics are trapped within the corresponding degenerate subspace, preventing relaxation to the true ground state. By introducing short imaginary-time steps, the \ifa{} scheme overcomes this obstruction, gradually reducing excited-state contributions and achieving high-fidelity ground-state preparation across all tested lattices. This hybrid real-imaginary-time framework therefore offers a robust approach to ground-state preparation in strongly correlated quantum systems.
The present work focuses on noiseless statevector simulations and small-to-intermediate system sizes, emphasizing the algorithmic behavior of \ifa{} rather than a detailed hardware-resource analysis. Practical implementations will require balancing the improved convergence provided by the ITE steps against their additional measurement and optimization costs. Extending the framework to noisy quantum hardware, larger lattices, and measurement-limited settings remains an important direction for future work.

\acknowledgments
This paper is supported by the Tohoku Initiative for Fostering Global Researchers for Interdisciplinary Sciences (TI-FRIS) of MEXT's Strategic Professional Development Program for Young Researchers.  L.N.T is funded by Vietnam National University, Ho Chi Minh City (VNU-HCM) under grant number B2026-18-18.
T.N.V.L. acknowledges helpful discussions with Vo Minh Phuoc regarding the VQE performance of the H$_4$ and H$_6$ molecules, and with Mai Nguyen Phuong Nhi on the FALQON method.

\section*{Data availability} 
The data that support the findings of this article are openly
available \cite{thanh2025}.

\appendix
\section{Jordan-Wigner mapping}
The Jordan-Wigner (JW) mapping converts fermionic operators into qubit operators.
Each qubit stores whether a fermionic orbital is empty or occupied:
\(|0\rangle\) = empty, \(|1\rangle\) = occupied.
To preserve the fermionic anti-commutation rule \(\{c_i,c_j^\dagger\}=\delta_{ij}\), every operator carries a string of \(Z\) gates that counts the parity of all fermions before it.
For a system with \(N\) qubits, the mapping is
\begin{align}
c_j &= Z^{\otimes (j-1)} \otimes \sigma^-_j \otimes I^{\otimes (N-j)},\\
c_j^\dagger &= Z^{\otimes (j-1)} \otimes \sigma^+_j \otimes I^{\otimes (N-j)},
\end{align}
where \(\sigma^\pm_j = \tfrac{1}{2}(X_j \pm iY_j)\).

For the number operator \(n_{i\sigma}=c_{i\sigma}^\dagger c_{i\sigma}\), the JW mapping gives
\(
n_{i\sigma} = \frac{I - Z_{i\sigma}}{2}.
\)
Therefore, the on-site interaction term becomes a diagonal qubit operator
\begin{equation}
n_{i\uparrow} n_{i\downarrow}
= \frac{1}{4}(I - Z_{i\uparrow})(I - Z_{i\downarrow}).
\end{equation}

The hopping term $c_i^\dagger c_j + c_j^\dagger c_i$ ($i < j$) involves the product of operators at different sites. The parity strings $Z$ cancel out on indices $k < i$ and $k > j$, leaving a string of $Z$ operators only on the qubits between $i$ and $j$
\begin{align}\notag
c_i^\dagger c_j + c_j^\dagger c_i
= \frac{1}{2}\big(
X_i Z_{i+1}\cdots Z_{j-1} X_j
+ Y_i Z_{i+1}\cdots Z_{j-1} Y_j
  \big).
  \end{align}
This parity string ensures that exchanging two fermions produces the correct minus sign.
Combining both parts, the mapped FH Hamiltonian reads
\begin{align}
\notag H_{\rm FH} = &-\frac{J}{2}\sum_{\langle i,j\rangle}
\big(X_i X_j + Y_i Y_j\big) Z_{i+1}\cdots Z_{j-1} \\ 
&+\frac{U}{4}
\sum_{\langle a,b\rangle} (I-Z_a)(I-Z_b),
\end{align}
where \(\la i,j\ra\) enumerates qubit indices corresponding to neighboring lattice sites (same spin) and \(\la a,b\ra\) denotes the qubits encoding spin-up and spin-down orbitals at the same site.

\bibliography{references}
\end{document}